\documentclass[aps,twocolumn,floatfix, showkeys, superscriptaddress, nofootinbib]{revtex4}
\usepackage{amsmath}
\usepackage{graphicx,bm,hhline}
\DeclareMathAlphabet{\pazocal}{OMS}{zplm}{m}{n}

\newcommand{\bcr}{{\bm R}}
\newcommand{\bk}{{\bm k}}

\newcommand{\ba}{{\bm a}}
\newcommand{\br}{{\bm r}}
\newcommand{\bx}{{\bm x}}
\newcommand{\bxp}{{\bx}^\prime}

\newcommand{\brp}{{\br}^\prime}

\newcommand{\hf}{\frac{1}{2}}

\newcommand{\wig}[1]{\mathrel{\hbox{\hbox to 0pt{\lower.6ex\hbox{$\sim$}\hss}\raise.4ex\hbox{$#1$}}}}

\newcommand{\lp}{l^\prime}
\newcommand{\mmp}{m^\prime}
\newcommand{\lpp}{l^{\prime\prime}}
\newcommand{\mpp}{m^{\prime\prime}}

\makeatletter
\renewcommand*\env@matrix[1][\arraystretch]{%
  \edef\arraystretch{#1}%
  \hskip -\arraycolsep
  \let\@ifnextchar\new@ifnextchar
  \array{*\c@MaxMatrixCols c}}
\makeatother

\begin{document}
\title{Multiple Scattering Theory for Dense Plasmas}

\author{C. E. Starrett}
\email{starrett@lanl.gov}
\affiliation{Los Alamos National Laboratory, P.O. Box 1663, Los Alamos, NM 87545, U.S.A.}

\author{N. Shaffer}
\affiliation{Los Alamos National Laboratory, P.O. Box 1663, Los Alamos, NM 87545, U.S.A.}

\date{\today}
\begin{abstract}
Dense plasmas occur in stars, giant planets and in inertial fusion experiments.  Accurate modeling of the electronic structure of these plasmas allows for prediction of material properties that can in turn be used to simulate these astrophysical objects and terrestrial experiments.  But modeling them remains a challenge.  Here we explore the Korringa-Kohn-Rostoker Green's function (KKR-GF) method for this purpose.  We find that it is able to predict equation of state in good agreement with other state of the art methods, where they are accurate and viable.  In addition, it is shown that the computational cost does not significantly change with temperature, in contrast with other approaches.  Moreover, the method does not use pseudopotentials -- core states are calculated self consistently.
We conclude that KKR-GF is a very promising method for dense plasma simulation.  
\end{abstract}
\pacs{ }
\maketitle

\section{Introduction}
In stars like our sun \cite{bailey15}, in white dwarfs \cite{Heinonen20}, and in inertial fusion experiments \cite{gomez14, lepape18} a state of matter variously known as warm and hot dense matter, or dense plasma, is reached.  To understand and model these physical systems it is necessary to be able to accurately predict the material properties, such as equation of state and transport properties, of these dense plasmas.  

There has been a serious and sustained effort to produce such accurate models.  For temperatures roughly less than the Fermi temperature, plane-wave based Kohn-Sham density functional theory molecular dynamics (DFT-MD) has become a widely used and reliable tool \cite{collins95, knudson18, hamel12, clerouin13a, hu14}.  To get to higher temperatures, orbital-free MD has proved useful \cite{lambert06a, danel15, kress11, karasiev14,profess08}, but is limited in its physical accuracy by the approximate kinetic-entropic functional used.  For example, we know of no orbital free functional that can predict discrete bound (or core) states \cite{sjostrom14,luo20}.  Very recently, stochastic DFT \cite{cytter18} has been shown to perform well at high temperature, but still uses pseudopotentials to represent core states, leading to transferability issues.  Moreover, stochastic DFT yields the electron density, not the Kohn-Sham orbitals, so standard linear-response theory for optical and transport coefficients can not be applied, and alternative methods such as time-dependent DFT will have to be developed \cite{baczewski16}.

Another high fidelity method is Path integral Monte Carlo (PIMC) \cite{militzer15,driver15a}, which is an excellent method for equation of state (EOS) at high temperature, but becomes prohibitively expensive at lower temperature due to the fermion sign problem.  To our knowledge, the method cannot directly produce other quantities of interest like opacity or transport coefficients.

More approximate methods include average atom models and their extensions to include ionic structure \cite{liberman, wilson06, starrett19, scaalp, perrot87, ovechkin16}.  These are DFT models of one fictitious averaged atom whose electronic structure represents an average over all atoms in the plasma.  These methods have been enormously useful, and still are the dominant method for production of large tables of data due to their reasonable fidelity and fast computation times.  However, due to the extreme difficulty in performing controlled EOS, opacity and transport coefficient measurements on dense plasmas in the laboratory, benchmarking these codes is largely done with reference to more complete models. 

In this paper we explore the Korringa-Kohn-Rostoker Green's function (KKR-GF) method, also known as the multiple scattering method, for calculating the electronic structure of dense plasmas \cite{korringa47, kohn54, ham61}.  KKR-GF is a general method for solving the Schr\"odinger or Dirac equation, but routinely, as here, it is used to solve the Khon-Sham(-Dirac) DFT equations \cite{ebert11, papanikolaou02, zeller11, zabloudil_book, mermin65}.  The method has been developed for many years in the solid state community \cite{zeller98,johnson84,thiess12, faulkner79}, and was introduced to the dense plasma community quite recently \cite{wilson11}, where an overview of the method and its potential benefits were presented, as well as some exploratory calculations.  Since then, some work has been done exploring the behavior of high temperature crystals, i.e. hot electrons with ions fixed in their crystal positions \cite{starrett18}, and some calculations of EOS, again using hot crystals have been presented \cite{grabowski20}.  However, the challenging question of the behavior of the method for disordered plasma environments has not, until now, been addressed.  Here we address this question, and present calculations for dense plasmas.  We find good agreement for EOS with existing state of the art methods, where they are accurate and tractable.  We find that the key to applying KKR-GF to these disordered environments lies in the concept of extra expansion centers.

In demonstrating that this method is viable for dense plasmas, we are introducing a method with a very high physical fidelity for calculating the electronic structure of dense plasmas.  It has the advantage of providing  an, in principle, exact solution of the Khon-Sham DFT equations; even core electrons are evaluated self-consistently, in contrast to pseudopotential-based methods.  The main weakness, as discussed later in this work, is the convergence of the multiple scattering sum.  The practical solution, used for many years when applying the method to low temperature systems, is demonstrated here to also give good results for dense plasmas.

\section{General theory}
The central quantity in the KKR-GF method is the one-electron Green's function $G(\br,\brp,z)$, defined as the solution to the Kohn-Sham inhomogeneous equation \cite{economou06}
\begin{equation}
\left[
z-H(\br)
\right]
G(\br, \brp,z)= \delta(\br-\brp)
\end{equation}
where $z$ is the (complex) energy and $H$ is the Kohn-Sham Hamiltonian.  The spectral representation of the Green's function for real energy $\epsilon$ is
\begin{equation}
G(\br, \brp,\epsilon)=
\sum_{\nu} \frac{\psi_\nu(\br) \psi_\nu^*(\brp)}{\epsilon-\epsilon_\nu +\imath \eta}
\end{equation}
where $\lim \eta \to +0$ should be taken, giving the so-called retarded Green's function, and the $\psi_\nu$ are the eigenstates with eigenvalues $\epsilon_\nu$.
Contact with the more familiar wave function formalism is made by taking the imaginary part of the Green's function 
\begin{equation}
\Im {G(\br, \brp,\epsilon)}=-\pi \sum_\nu
\psi_\nu(\br) \psi_\nu^*(\brp) \delta(\epsilon-\epsilon_\nu)
\label{img}
\end{equation}
The electron density is given by
\begin{equation}
n_e(\br) = -\frac{2}{\pi}\Im \int_{-\infty}^{\infty}d\epsilon f(\epsilon,\mu)G(\br,\br,\epsilon)
\label{ne_gf}
\end{equation}
where the $2$ is due to spin degeneracy, the integral is along the real energy axis, and $f$ is the Fermi-Dirac function with chemical potential $\mu$.  Using equation (\ref{img}), equation (\ref{ne_gf}) clearly recovers the usual expression for $n_e(\br)$.
\begin{figure}
\begin{center}
\begin{tabular}{c}
\includegraphics[scale=0.3]{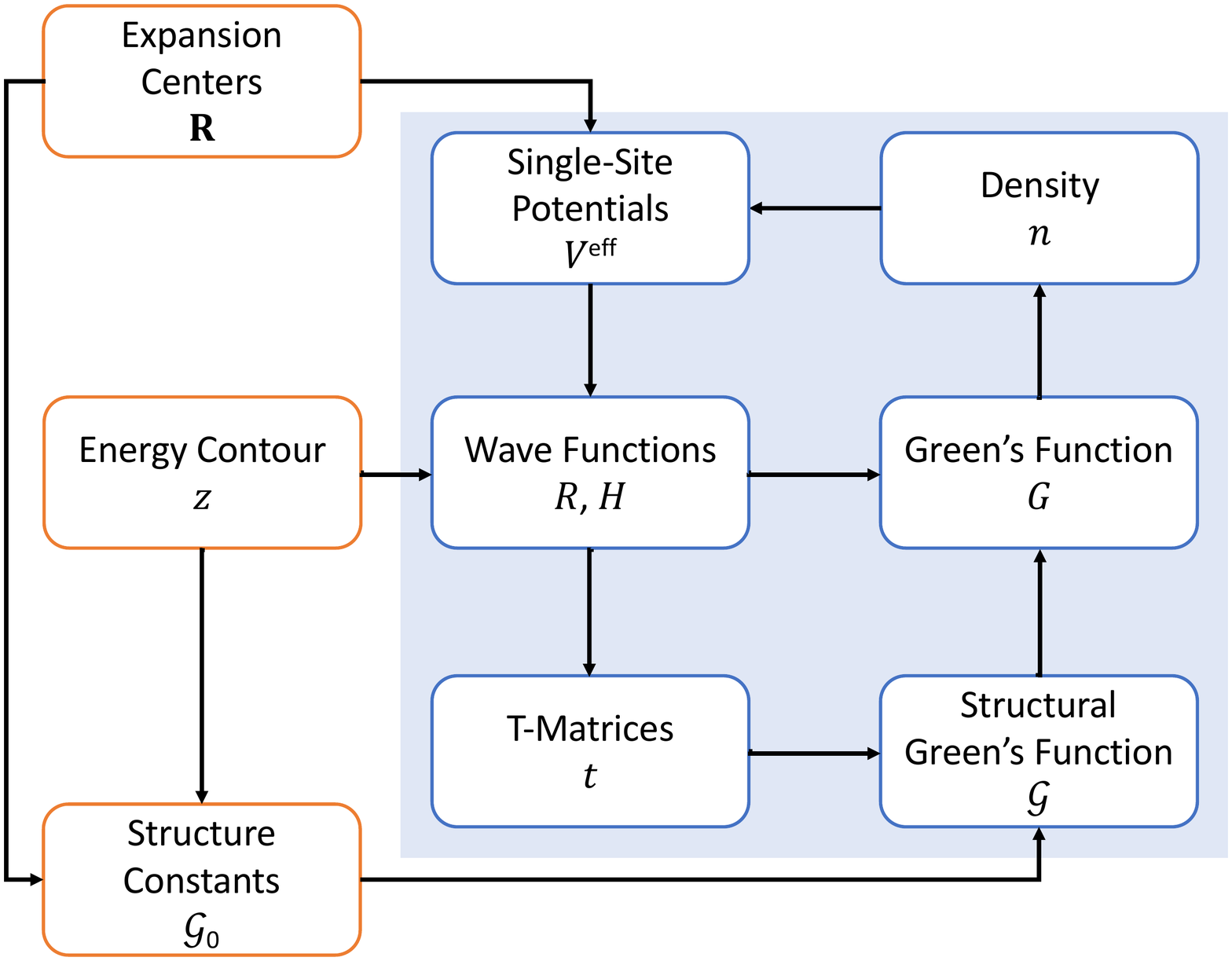} 
\end{tabular}
\end{center}
\caption{(Color online) Flow chart describing the computational workflow.  Inside the shaded (blue) region is the self-consistent field (SCF) procedure, outside the shaded area, in orange boxes, are one-off steps that are decided or calculated before the SCF procedure starts.}
\label{fig_fc}
\end{figure}

An important concept in the KKR-GF \cite{korringa47, kohn54, ham61} method is the multi-site expansion of the Green's function, where the spatial dependence of $G(\br,\brp,z)$ is expanded about a set of fixed centers.  These centers are used to define cells that are non-overlapping and space filling.   Starting with the known free electron Green's function 
\begin{equation}
g(\bx ,\bxp,z)=
-\frac{m}{ 2\pi } \frac{\exp(\imath p|\bx-\bxp|)}{|\bx-\bxp|}
\end{equation}
where $p=\sqrt{2m z}$, one expands the Green's function about a set of centers $\{\bcr ^n\}$ using spherical harmonics $Y_{lm}(\hat{\br})$.  Note that we are using atomic units in which $\hbar = a_B = 4\pi \epsilon_0 =e=1$, but we leave the electron mass $m$ dependence as it allows easy conversion between Rydberg and Hartree atomic units.
Letting $\bx=\br+\bcr^n$ and  $\bxp=\brp+\bcr^{n'}$, we arrive at
\begin{equation}
\begin{split}
g(\br+\bcr^n ,\brp+\bcr^{n'},z)=&
\delta_{nn'} 2m
\sum_{L=0}^\infty h^\times_L(\br_>,z) j_L(\br_<,z)\\
&\!\!\!\!\!\!\!\!\!\!\!\!
+2m\sum_{LL'}^\infty j_L(\br,z)
{\cal G}^{nn'}_{0,LL'}(z)   
j_{L'}^\times(\brp,z)
\end{split}
\label{gfexp}
\end{equation}
where $L=\{l,m\}$, $j_L(\br,z) = j_l(p r) Y_{l,m}(\hat{\br})$, $h_L(\br,z) =-\imath p h_l(p r) Y_{l,m}(\hat{\br})$, $j_l$ ($h_l$) is the spherical Bessel (Hankel) function, and ${\cal G}^{nn'}_{0,LL'}$ are the structure constants (appendix \ref{app_struct}).  The superscript $\times$ indicates that the complex conjugate of the spherical harmonic should be taken.  Also, $\br_>$ ($\br_<$) means that one should take the vector with greater (lesser) magnitude of $\br$ and $\brp$.

The expression (\ref{gfexp}) can be generalized to the case of non-free electrons, giving
\begin{equation}
\begin{split}
G(\br+\bcr^n ,\brp+\bcr^{n'},z)=&
\delta_{nn'}2m
\sum_{L=0}^\infty H^{n,\times}_L(\br_>,z) R^{n}_L(\br_<,z)\\
&\!\!\!\!\!\!\!\!\!\!\!\!
+2m
\sum_{LL'}^\infty R^{n}_L(\br,z)
{\cal G}^{nn'}_{LL'}(z)   
R^{n'\times}_{L'}(\brp,z)
\end{split}
\label{gfexp2}
\end{equation}
where $R^n_{L}$ is the regular solution to the Kohn-Sham equation in cell $n$ and $H^n_L$ is  the irregular solution, and ${\cal G}^{nn'}_{LL'}$ are the structural Green's function matrix elements (appendix \ref{app_struct}).  
The double sum term in equation (\ref{gfexp2}) can be thought of as a multisite correction the the single site Green's function, defined by the single summation term.  This so-called `multiple scattering' term corrects the boundary condition applied to the single site solution (i.e. the free electron boundary condition), replacing it with the correct form so that all incoming and outgoing waves are matched at the cell boundaries.  Another way to think of it, is that the multiple scattering term introduces quantum diffraction, where electron scattering from multiple sites can interfere.

The structural Green's function matrix $\bm{\mathcal G}$  is related to the structure constants matrix $\bm{\mathcal G}_0 $ and to the so-called $\bm t$-matrix, which encapsulates the scattering properties of all the cells, by Dyson's equation
\begin{equation}
{\bm{\mathcal G}}(z)=
{\bm{\mathcal G}}_0(z)
\left(I-
{\bm{t}}(z)
{\bm{\mathcal G}}_0(z)
\right)^{-1}
\end{equation}
This expression is known as the fundamental equation of multiple scattering theory.  It is solved by matrix inversion.  Each term is a matrix indexed by the site indices $n$ and $n'$, and by $L$ and $L'$.  The $\bm t$-matrix is site diagonal.

\begin{figure}
\begin{center}
\begin{tabular}{c}
\includegraphics[scale=0.35]{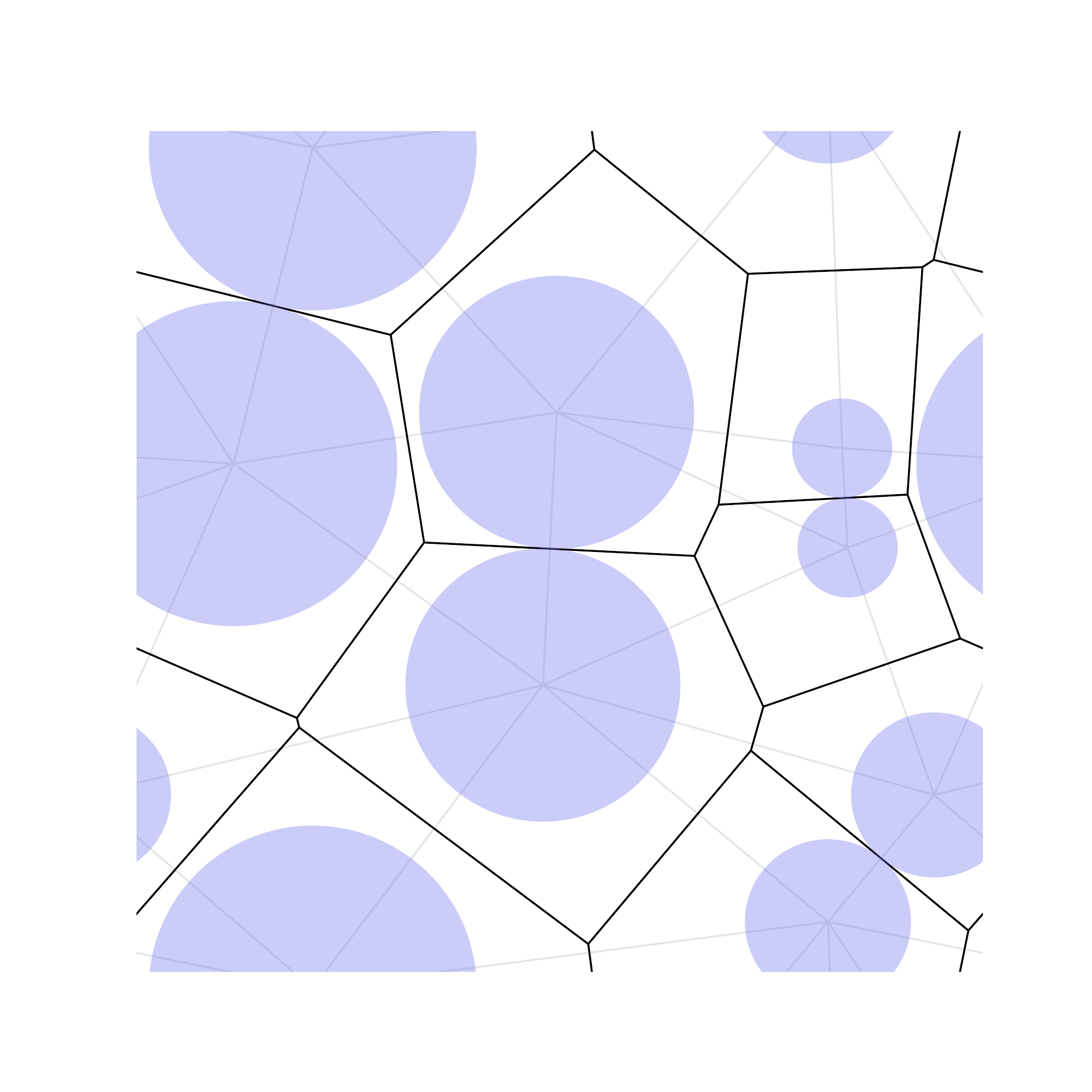} \\[0.2cm]
\includegraphics[scale=0.35]{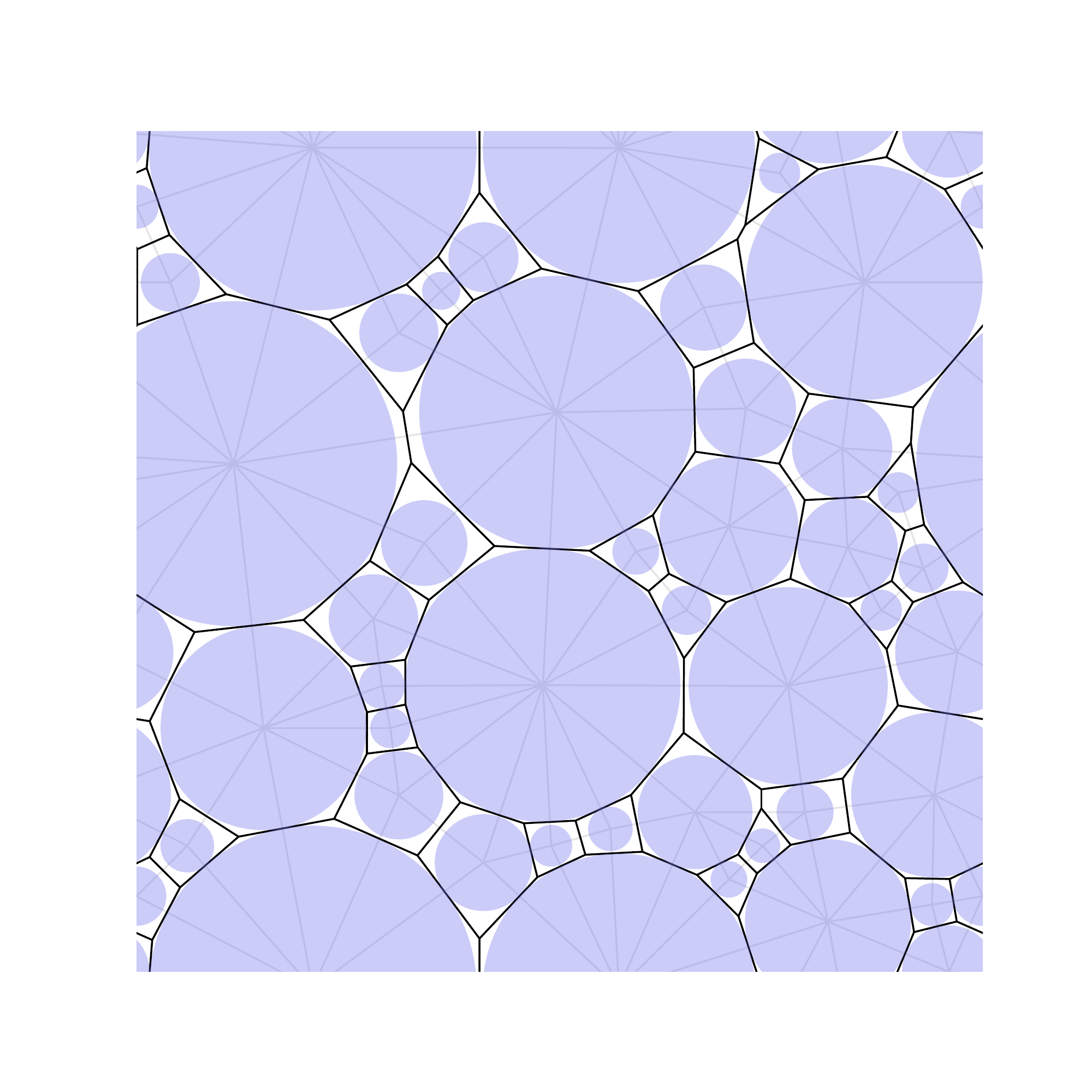} 
\end{tabular}
\end{center}
\caption{(Color online) Schematic example of power tessellation. Top panel shows the 2-D tessellation for 8 centers, which would correspond to the nuclear positions in 3-D.  The circles correspond to the muffin tin spheres, and the black line demarcate the cells.  In the bottom panel 35 extra centers have been added in such a way as to minimize the area (volume) not in the muffin tins.  Notice that the power tessellation preserves the original 8 spheres as more centers are added.}
\label{fig_ec}
\end{figure}

\section{practical considerations}
In this section we give details on the implementation of this method.  A flow chart summary of the method is given in figure \ref{fig_fc}.
At a high level, the method involves dividing space into non-overlapping, space-filling cells, solving the Kohn-Sham equations within each cell independently, and combining these solutions via Dyson's equation to form the Green's function.  In turn this Green's function is used to obtain a new electron density to feed back into the Kohn-Sham equations  until self-consistency is achieved.

\subsection{Tessellation of Space}
To implement the KKR-GF method in practice the first step is to tessellate space.  For regular crystals a Voronoi tessellation has been used with success \cite{ebert11}.  For crystals with more than one atomic species, using a power tessellation has been shown to be numerically advantageous \cite{alam09}.  In principle the Green's function should be independent of the tessellation, but to minimize the number of spherical harmonic expansion terms, ideally the expansion regions (i.e. the cells) should  be as spherically symmetric as possible.  This is why the power tessellation is better for mixtures; it allows one to weight cell sizes, in contrast to the Voronoi procedure which tessellates purely on the positions of the expansion centers.  
In the implementation applied here we have used the power-tessellation code of reference \cite{bernal}. 

\begin{figure}
\begin{center}
\begin{tabular}{c}
\includegraphics[scale=0.5]{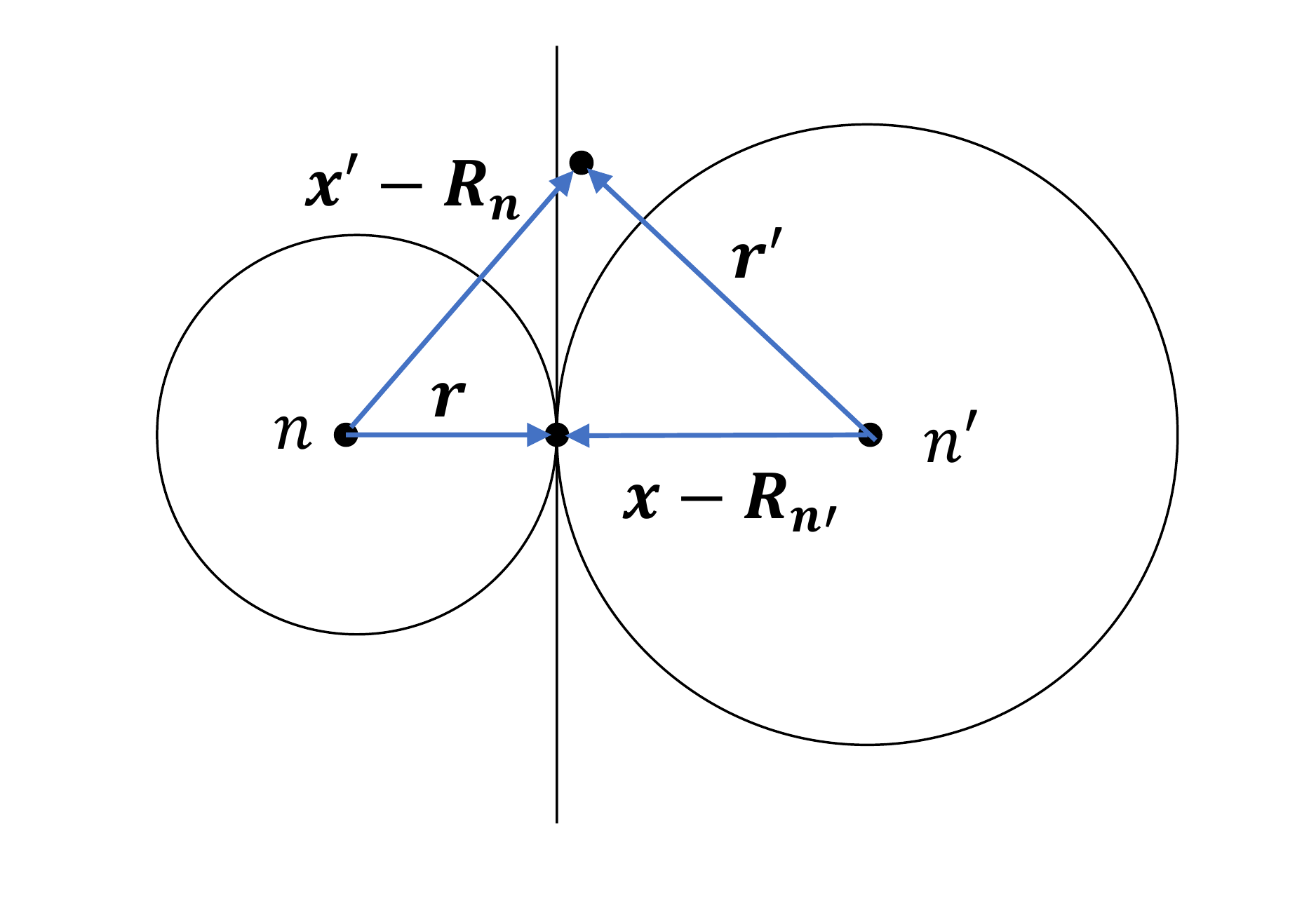} 
\end{tabular}
\end{center}
\caption{(Color online) Example of geometry in which one of the inequalities is violated.  The circles represent the muffin tin spheres and the vertical line is the boundary plane.   For the points $\br$ and $\brp$, the inequality  $|\brp|<|\bx-\bcr^{n'}|$ is violated, while the others are satisfied.}
\label{fig_geo}
\end{figure}

There is another important consideration when making this tessellation: the derivation of equations (\ref{gfexp}) and (\ref{gfexp2}) rely the following sets of inequalities being satisfied \cite{morgan77, zeller15}:
\begin{equation}
|\br|<|\bcr^n-\bcr^{n'}| \mbox {  and  } |\brp|<|\bx-\bcr^{n'}|
\label{cond1}
\end{equation}
and
\begin{equation}
 |\brp|<|\bcr^{n'}-\bcr^{n}|
\mbox{  and  } 
 |\br|<|\bxp-\bcr^n|  
\label{cond2}
\end{equation}
These are satisfied for all combinations of $\br$ and $\brp$ if the cells $n$ and $n'$ are the largest possible, non-overlapping spheres for a given set of centers; these spheres are called the muffin tin spheres.  For space filling tessellations the left hand set of inequalities can always be made to be satisfied by adding extra centers (see figure \ref{fig_ec}).  
For power tessellations, one of the right-hand inequalities can also always be satisfied \cite{morgan77}. However, there will always be certain pairs of points where the other right-hand inequality is violated (figure \ref{fig_geo}).  This leads to slow or even conditional convergence of the double sum in the expansion (\ref{gfexp2}). 

The practical solution to this conditional convergence issue is to truncate the double summation to $l,l'\leq l_{max}^{ms}$ \cite{zeller15,gonis12}, where $l_{max}^{ms}=2,3, \mbox{ or } 4$.  This works because using $l_{max}^{ms}=2,3,\mbox{ or }4$ covers the orbital angular momenta of the chemically relevant electrons and bound states.  Higher $l$ states correspond to higher energy electron states that are therefore more like free electrons which are less affected by multisite contributions.   It is important to note that adding higher order terms in the expansion (larger $l_{max}^{ms}$) may actually lead to worse results due to this conditional convergence \cite{zeller15,morgan77}.  This is the key weakness of the method.  It can be systematically ameliorated by increasing the number of expansion centers.  However, as we shall see in section \ref{sec_sen}, relative insensitivity of the equation of state to the choice of $l_{max}$ within the range 2 to 4, affirms the validity of this practical solution.

Once the expansion centers have been decided the structural constants $\bm{\mathcal G}_0$ are calculated.  These depend only on the positions of the centers and the energies needed (see later).  Hence for a given set of centers, they need only be calculated once.  We have assumed a periodic supercell in our calculations.  This artificial construct is widely used for simulations of disordered systems.  The corresponding structural constants are thus calculated in Fourier-space and transformed back to real-space, see appendix \ref{app_struct}.  An alternative method that we do not explore here, is to neglect long range effects on the Green's function, and evaluate the structure constants using a finite cluster of atoms \cite{ebert11,rehr10,wang95}.

\subsection{Single-Site Solutions}
The next step is to solve the single-site problem in each cell.  This means solving the Kohn-Sham equation for the regular $R^n_{L}(\br)$ and irregular $H^n_{L}(\br)$ solutions, as well as the ${\bm t}$-matrix for each site ${\bm t}^n$.  This problem is considerably simplified if the effective potential inside the cell $V^{eff,n}(\br)$ is  spherically symmetric.   In cells around a nucleus, the electron density (and effective potential) will be approximately spherically symmetric due to the dominance of the nuclear potential.  In cells not containing a nucleus, the electron density will be reasonably close to that of a free electron gas, which is uniform.  Hence for this initial evaluation of the method for disordered dense plasmas, it is reasonable to adopt this spherically symmetric approximation for $V^{eff,n}(\br)$ within the cells.  This approximation is called the muffin tin approximation.

In the muffin tin approximation the effective potential inside the muffin tin spheres is spherically averaged:
\begin{equation}
V^{eff,n}_{mt}(r) = \frac{1}{4\pi} \int d\hat{\br} V^{eff,n}(\br) \mbox{ for } r < R_{mt}^n
\end{equation}
where $R_{mt}^n$ is the muffin tin radius for cell $n$.
Outside the muffin tin spheres, in the interstitial ($is$) region, the potential takes an average value -- the muffin tin constant.  This is determined by 
\begin{equation}
\bar{V}^{mt} = \sum_n \frac{1}{V_{is}^n}   \int_{\br \in {is}} d^3r V^{eff,n}(\br)
\end{equation}
where the sum is over all cells, and $V_{is}^s$ is the volume of the interstitial region for cell $n$.  Hence the potential inside each cell is spherically symmetric.  

An important point worth clarifying is that
even though $V^{eff,n}_{mt}(r)$ is spherically symmetric, the Green's function and electron density inside a cell are not, in general.  The symmetry is broken by the multi-site term in the Green's function.  This results from the coupling of different spherical harmonics in the double sum term in equation (\ref{gfexp2}).  

The regular and irregular solution must be matched to their free electron forms at the edge of the cell
\begin{eqnarray}
R_L^n(\br,z) & = & \left(  j_l(pr)- \imath p h_l(pr) t^n_l(p) \right) Y_{lm}(\hat{\br}) \\
H_L^n(\br,z) & = &   -\imath p h_l(pr) Y_{lm}(\hat{\br})
\end{eqnarray}
where the ${\bm t}^n$-matrix is now diagonal in the muffin tin approximation.  The regular solution is found by integrating out from the origin, and the irregular solution is found by integrating inward \cite{starrett15}.  In matching the numerical solutions to these analytic forms the ${\bm t}^n$-matrix is found.

Going beyond the MT approximation is the goal of so-called ``full potential'' calculations  \cite{zeller98,asato99,ogura05}. This introduces significant complexity to the single-site problem. For example, the ${\bm t}^n$-matrix ceases to be diagonal in $L$. The importance of full-potential versus MT for disordered plasma applications will be considered in the future.

\subsection{Efficient Evaluation of Multi-Site Effects}
With the single-site problem solved and the structural Green's function evaluated, one can construct the Green's function, equation (\ref{gfexp2}).
The multiple scattering term is expensive to evaluate due in part to the double sum (equation (\ref{gfexp2})) but also because 
the solution of Dyson's equation for the structural Green's function matrix involves a matrix inversion for each energy and k-space integration point (appendix \ref{app_struct}).  The size of this 
matrix is $N(l_{max}^{ms}+1)^2$ where $N$ is the number of expansion centers and $l_{max}^{ms}$ is the maximum $l$ for both summations.  Since dense matrix inversion cost scales as size of the matrix cubed, it is clearly desirable to keep $N$ and $l_{max}^{ms}$ as small as possible 
\footnote{Note, however, that there exist so-called `screening transform' methods which sparsify the matrix inversion and permit more favorable scaling for large systems \cite{szunyogh94}.}.   As noted above, only the first few terms are included in the double sum ($l_{max}^{ms}=2,3$ or $4$), which serves to limit computational expense.  As explored in reference \cite{starrett18}, higher $l$ terms are kept in the single site part of the Green's function.  This is expected to be a good approximation because higher $l$ states correspond to higher energy states, and hence are more free-electron like, leading to small multiple scattering contributions.

Another way to reduce computational expense is to evaluate the multi-site term only in the energy range where it is significant.  We expect tightly bound states to be unaffected by multi-site interactions due to their localized nature.  At the other extreme, electrons in high-energy states should be insensitive to the details of multi-site interactions due to their free electron-like character.  Hence we introduce a minimum $E_{min}$ and maximum energy $E_{max}^{ms}$ for  multi-site contributions, outside of which only the single-site term contributes.

Even if the multiple scattering Green's function were evaluated only in the range $E_{min}$ to $E^{ms}_{max}$, it would be a rapidly varying function of the energy there, requiring many integration points to accurately resolve the multiple scattering contribution to the density, equation (\ref{ne_gf}). This is avoided by deforming the energy integration path into the upper half of the complex plane.
This technique results in
much smoother integrands, fewer integration points, and most importantly avoids the poles on the real axis, which correspond to the eigenstates of the system.  Hence, one does not actually find these eigenstates, nor are they needed.  The (retarded) Green's function has no poles in the upper half complex plane.  However, the Fermi-Dirac function does have poles, at the so-called Matsubara frequencies: $z=\mu +\imath \pi (2j-1)k_B T$, for $j=1,2,\ldots$.  Their residues can be evaluated directly, or avoided by choosing a contour with imaginary part of energy everywhere less than $\pi k_B T$.  Note also, that the poles are avoided if $\mu < E_{min}$ or $\mu > E_{max}^{ms}$.  

\begin{figure}
\begin{center}
\includegraphics[scale=0.3]{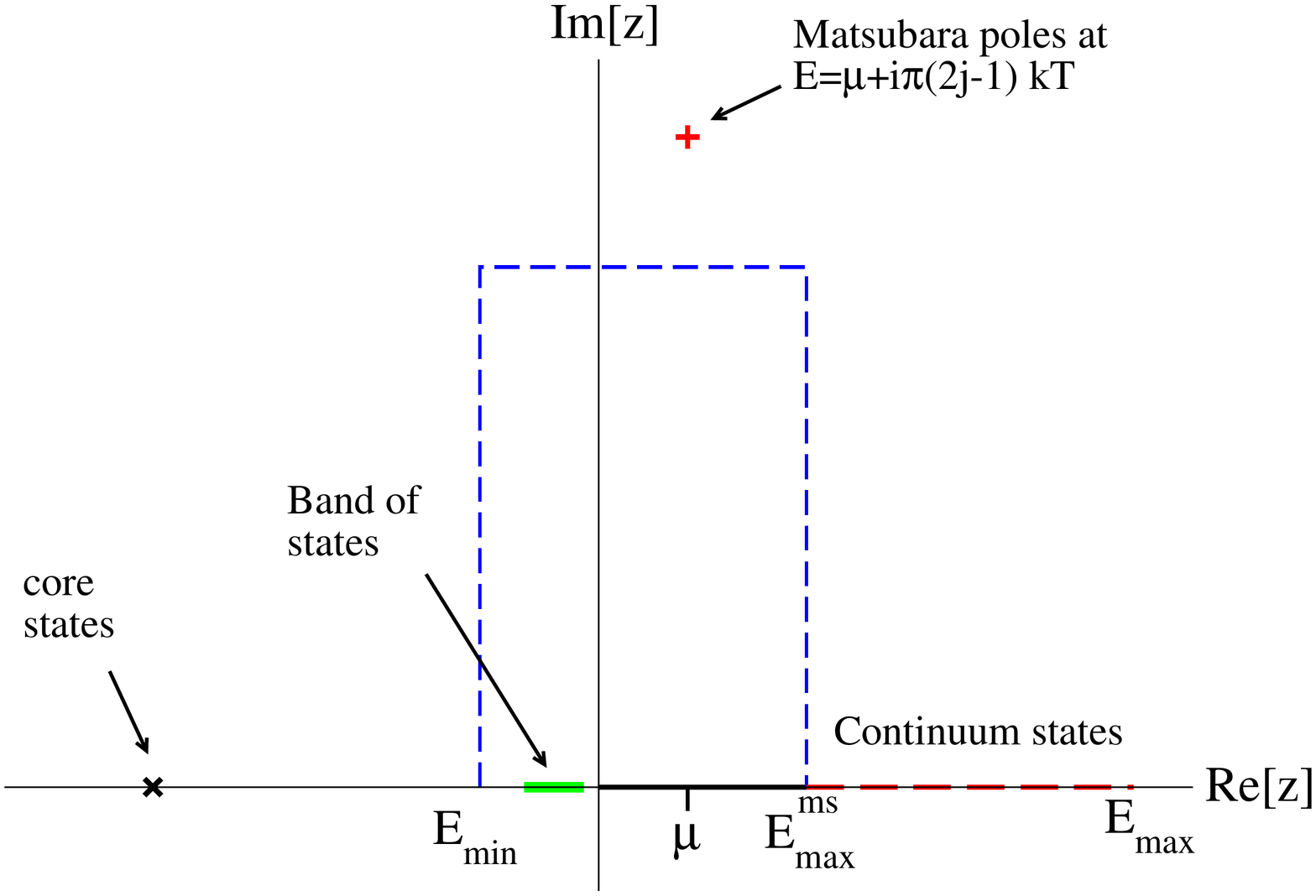}
\end{center}
\caption{(Color online) Schematic of the different types of states that occur on the real energy axis.  For large negative energy only discrete core states exist, and one only needs the numerically inexpensive single-site solution.  Between $E_{min}$ and $E_{max}^{ms}$ valence states exit, which are strongly affected by multi-site effects.  In this region the full Green's function expression is evaluated for $l \leq l_{max}^{ms}$.  For $l>l_{max}^{ms}$ only the single-site contribution is needed.  For energies greater than $E_{max}^{ms}$, again only the single site solution is needed.}
\label{fig_con}
\end{figure}

This strategy is summarized in figure \ref{fig_con}.  For large negative energy only the single-site term is evaluated, and and any core states that exist are accounted for using equation (\ref{img}).  Between $E_{min}$ and $E_{max}^{ms}$ the full Green's function expression is evaluated using contour integration.  Note that, for this region $l_{max}^{ss}$ for the single-site term is as large as needed for convergence, whereas $l_{max}^{ms}$ for the multi-site term is fixed at a small values (e.g. 2, 3, or 4).  Above $E_{max}^{ms}$ only the single site term is evaluated, up to some maximum energy $E_{max}$, determined by integrand having become negligibly small.

With the Green's function determined, the electron density can be evaluated, and the usual self consistent field procedure is followed.
The Poisson equation is solved using the technique presented in \cite{starrett17}.
The self-consistent field problem is solved using Eyert's method \cite{eyert96}, and
normally takes less than 10 iterations to converge, using the Thomas-Fermi cell model
to generate an initial guess \cite{feynman49}.  For spatial integrations over the cells, we use the isoparametric integration method of reference \cite{alam11a}.  Note also, that in the calculations presented here we use the
temperature dependent local density approximation \cite{ksdt}.

\subsection{Thermodynamics}
To study the convergence properties of the KKR-GF method in disordered plasmas, we focus on thermodynamic properties. These require the electron density of states in addition to the density.
The density of states per volume $V$ is also found from the Green's function
\begin{equation}
\chi(\epsilon)=  -\frac{2}{\pi}\Im \int_{V}d^3r\, G(\br,\br,\epsilon)
\label{dos_gf}
\end{equation}
The pressure is given by the Virial expression
\begin{equation}
P = \frac{1}{V} \left[ 
\frac{U^k + F^{el}}{3} 
\right] + P^{xc} + P^I
\label{pvir}
\end{equation}
where $F^{el}$ is the electrostatic free energy \cite{starrett17},  $P^I$ is the ideal ion pressure,
and $U^k$ is the internal kinetic energy
\begin{equation}
U^k  = 
\int_{-\infty}^{\infty}d\epsilon f(\epsilon,\mu)\chi(\epsilon)\epsilon - 
            \int_{V}d^3r V^{eff}(\br) n_{e}({\br}).
\end{equation}
and assuming the local density approximation
\begin{equation}
P^{xc}= \frac{1}{V}  \left[ -F^{xc} + \int_{V}d^3r\, n_{e}(\br) V^{xc}(\br)\right]
\end{equation}
where $F^{xc}$ is the exchange and correlation free energy, $V^{xc} = \delta F^{xc}/\delta n_e$.

Evaluating the plasma equation of state also requires averaging over an ensemble of realistic ion positions.  We generate these using the PAMD (pseudo-atom molecular dynamics) model \cite{starrett15a}.  This model uses a DFT average atom calculation to produce a force field for use in classical molecular dynamics.  It gives an accurate set of ion transport coefficients and pair distribution functions for dense plasmas \cite{starrett14,daligault16}.  It is numerically inexpensive and we use it to generate sets of ion positions that are well separated in time and hence, are uncorrelated.  This is a reliable and accurate procedure for dense plasmas.  For lower temperature liquids and warm dense matter with transient chemical bonds, PAMD will not produce accurate ion positions.  In principle, it is possible to calculate forces on ions directly from the KKR-GF solution \cite{zeller11_book}.  However, we have not attempted that yet and we note that results shown are firmly in the regime where PAMD is accurate.

\section{Results}
\subsection{The Effect of Extra Expansion Centers}
\begin{figure}
\begin{center}
\includegraphics[scale=0.4]{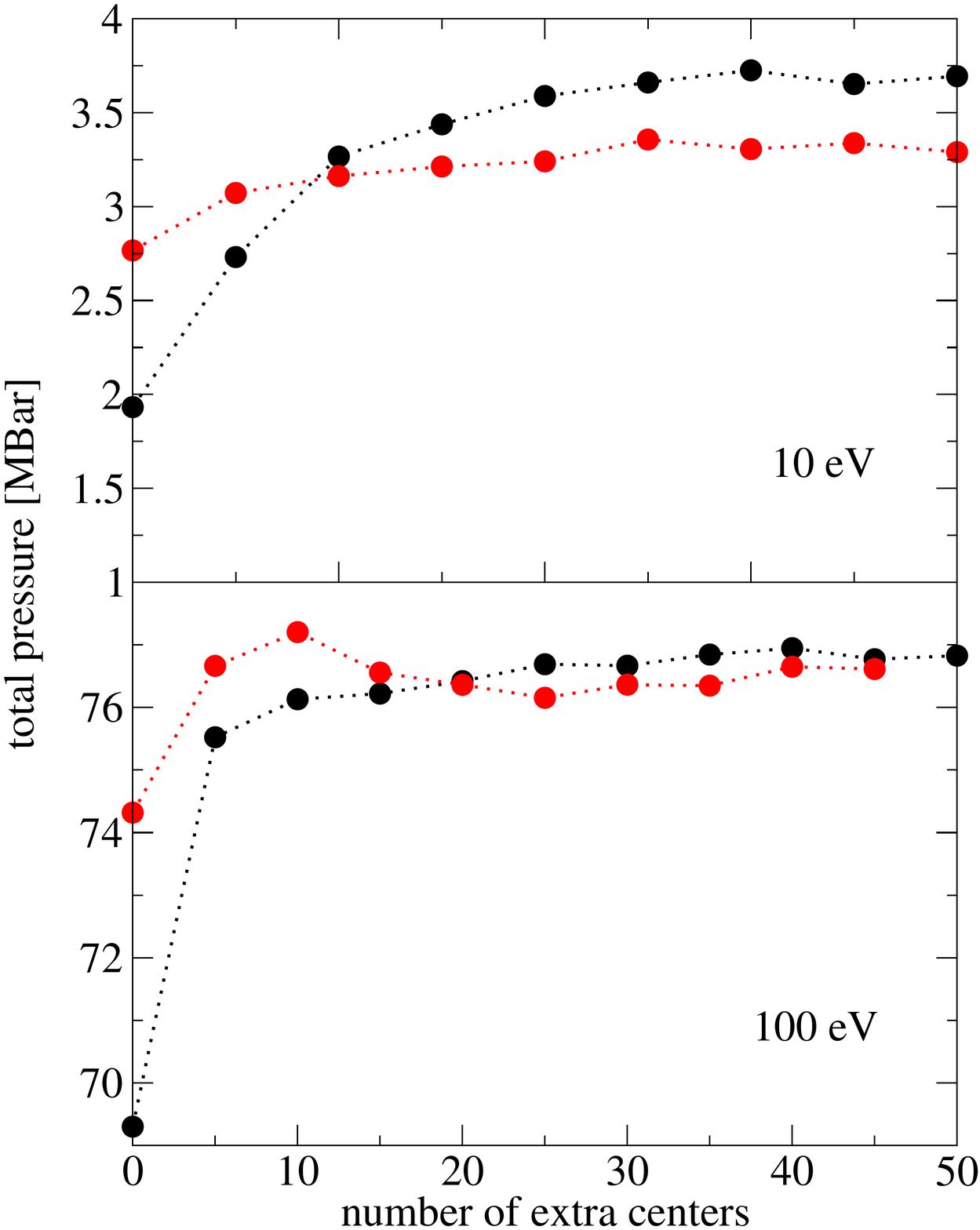}
\end{center}
\caption{(Color online) Convergence of pressure with the number of extra centers for 8 aluminum atoms, at solid density and temperatures of 10 eV (top panel) and 100 eV (bottom panel).  The dotted lines
serve as a guide to the eye.  The two lines correspond to two sets of ion configurations (i.e. positions).  The results are for $l_{max}^{ms}=2$.
}
\label{fig_ns}
\end{figure}

\begin{figure}
\begin{center}
\begin{tabular}{c}
\includegraphics[scale=0.4]{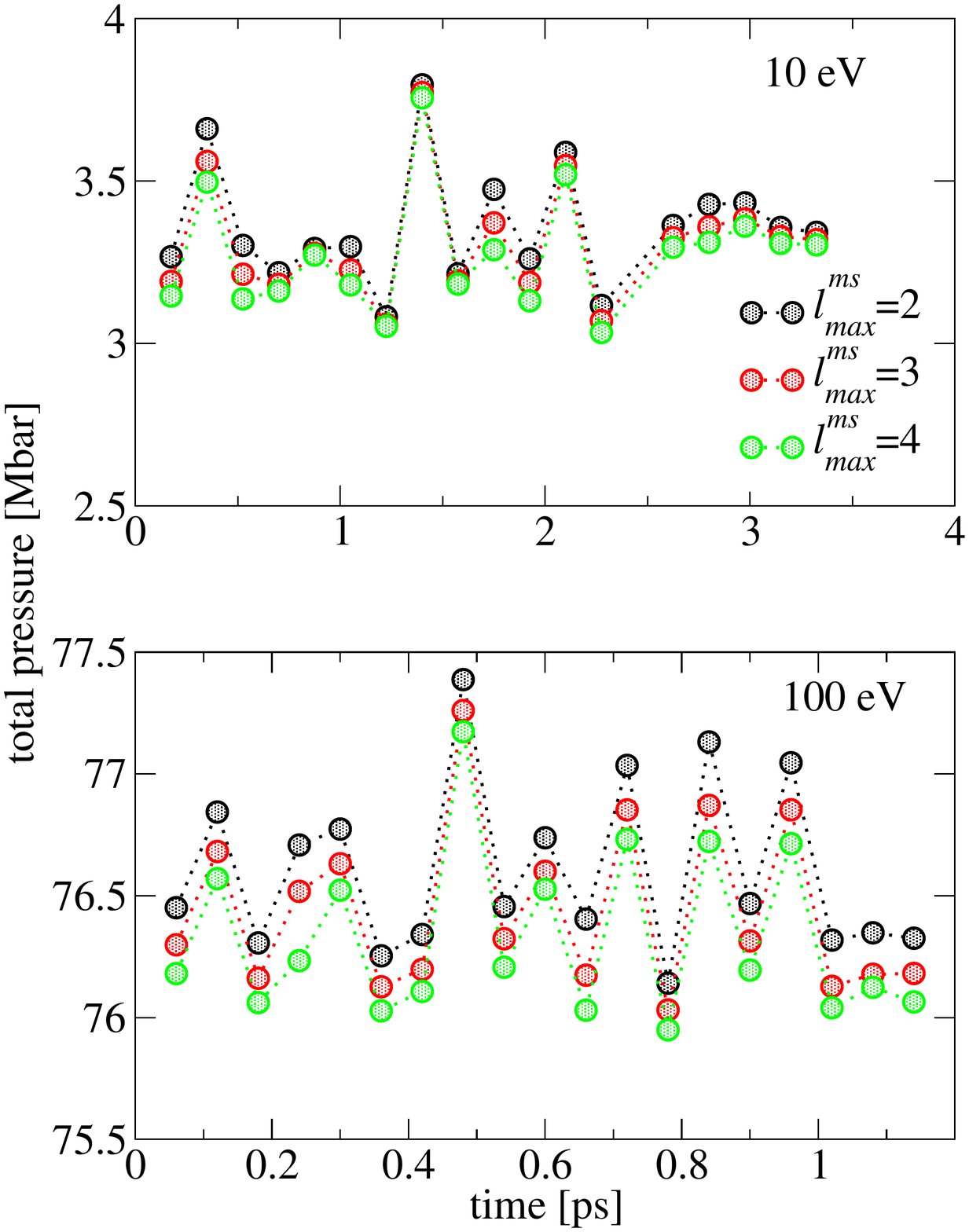}
\end{tabular}
\end{center}
\caption{(Color online) Total pressure for solid density aluminum at 10 eV (top panel) and 100 eV (bottom panel), versus simulated time.  Shown are results for $l_{max}^{ms}=2,3,\mbox{ and } 4$.  These results use 35 extra centers.
}
\label{fig_ptot}
\end{figure}

In figure \ref{fig_ns} we show the total pressure as a function of the number of extra centers for two different temperatures, for aluminum at solid density.  For each temperature, we show results for two different sets of ion positions.  We are using 8 nuclei in the simulation volume.   The number of extra centers corresponds to the number of cells not containing a nucleus.  We see that on adding centers the pressure initially changes rapidly and then settles down to a rough constant.  There are still some fluctuations in the pressure even after the value settles down.  The pressure initially changes rapidly because the inequities (\ref{cond1}) are strongly violated.  As extra centers are added, the expansion becomes valid and the value settles down.  However, as we are using the muffin-tin approximation, adding extra centers treats the electrons in the interstitial region slightly differently to before, hence the fluctuations.   At the same time, the region being modeled with the conditionally convergent double summation also reduces in size.  The exact interplay of these errors is not yet known but could be examined in detail with a future full-potential treatment.

It is also noted that the 100 eV case reaches its (roughly) constant pressure more quickly than the 10 eV case.  We can understand this behavior by noting that at high temperature the multiple scattering effect is of less relative importance as there are more high energy electrons that are insensitive to multi-site effects.

\subsection{Sensitivity to Multiple Scattering Summation \label{sec_sen}}
In figure \ref{fig_ptot} we shown the total pressure as a function of simulation time for the same 10 and 100 eV aluminum plasmas.   Results are shown for $l_{max}^{ms}=2, 3, \mbox{ and } 4$.   The average pressures and standard deviations are (in Mbar)
\begin{center}
\def\arraystretch{1.5}%
\begin{tabular}{c cc}
$l_{max}^{ms}$ & 10 eV & 100 eV \\
\hline
2       & 3.36 $\pm$ 0.18 & 76.6 $\pm$ 0.35 \\
3       & 3.31 $\pm$ 0.18 & 76.4 $\pm$ 0.34 \\
4       & 3.28 $\pm$ 0.18 & 76.3 $\pm$ 0.33
\end{tabular}
\end{center}
where we give results at temperatures of 10 and 100 eV, for $l_{max}^{ms}=2, 3, \mbox{ and } 4$.  The pressure does not change strongly with $l_{max}^{ms}$, indicating that the strategy of using a physically relevant $l_{max}^{ms}$ is meaningful.   Comparing the KKR-GF approach with other methods we have
\begin{center}
\def\arraystretch{1.5}%
\begin{tabular}{c c c c c}
T [eV] & KKR-GF & DFT-MD &  PAMD  &  {\texttt Tartarus} \\
\hline
10       & 3.31 $\pm$ 0.18 & 3.34  &  3.19  & 2.93 \\
100     & 76.4 $\pm$ 0.34 & --      &   76.3  & 75.4
\end{tabular}
\end{center}
where we have used $l_{max}^{ms}=3$.  At 10 eV we compare with DFT-MD results \cite{kress_priv} and find agreement within the error bars.  We also compare with PAMD \cite{starrett16} which also gives reasonable agreement, and the average atom model {\texttt Tartarus} \cite{starrett19}, which gives too low a pressure.   Since PAMD is based on the idea of correcting average atom pressures for ionic disorder, the difference between them implies that the lack of this effect in the {\texttt Tartarus} model is the cause of the difference.   At 100 eV the relative change with $l_{max}^{ms}$ is smaller than at 10 eV.  This is due to the decreased relative importance of multi-site effects, as noted earlier.  Agreement with PAMD is again good.   

The rather large fluctuations in the pressure seen in figure \ref{fig_ptot} are due to the small number of nuclei in the simulation, i.e. eight nuclei.  There is no restriction of the method to small numbers of atoms.  Indeed, linear scaling variations exist that could be exploited for plasma applications \cite{zeller11_book, wang95, thiess12}.  However, our results are enough to demonstrate the viability of the method, our main aim.

\begin{table}[]
\begin{center}
\bgroup
\def\arraystretch{1.5}%
\begin{tabular}{c cccc}
T [eV] & P [MBar]        &  Time [mins] & ${\cal G}_0$ time [mins]  \\
\hline
10       & 3.36 (0.182)  &  60               & 50 \\
20       & 7.56 (0.167)  &  61               & 49 \\
50       & 27.9 (0.473)  & 59                & 46 \\
100     & 76.44 (0.351) & 64               & 54 \\
200    & 194.6 (1.13)  & 61                 & 53 \\
500    &  600.5 (1.23) & 61                 & 54 \\
1000  &  1306 (3.20)  & 63                 & 53 
\end{tabular}
\egroup
\end{center}
\caption {Total pressure and wall time for solid density aluminum across two orders of magnitude of temperature.  Results are for $l_{max}^{ms}=2$, with 8 atoms and 35 extra spheres.   Wall time is the time taken to converge for a single set of ion positions using a single node with 18 threads for all cases.  ${\cal G}_0$ time is the wall time for the calculation of the structural Green's function.  Clearly, the results demonstrate an independence of wall time on temperature.  \label{tab_time}}
\end{table}
Table \ref{tab_time} shows total pressure along the 2.7 g/cm$^3$ isochore from 10 to 1000 eV and the wall time for one complete self consistent field cycle (to convergence) for a given set of ion positions.  The key point is that, in practice, the wall-time for KKR-GF method does not significantly change with temperature, in stark contrast with plane wave DFT-MD \cite{sjostrom14}.  Also shown is the wall-time for calculation of the structural constants, which takes roughly 5/6 of the total execution time.  This can probably be dramatically improved, either through algorithmic improvements, exploitation of modern parallelism, or switching to real-space approximations \cite{rehr10, wang95}.

\begin{figure}
\begin{center}
\includegraphics[scale=0.4]{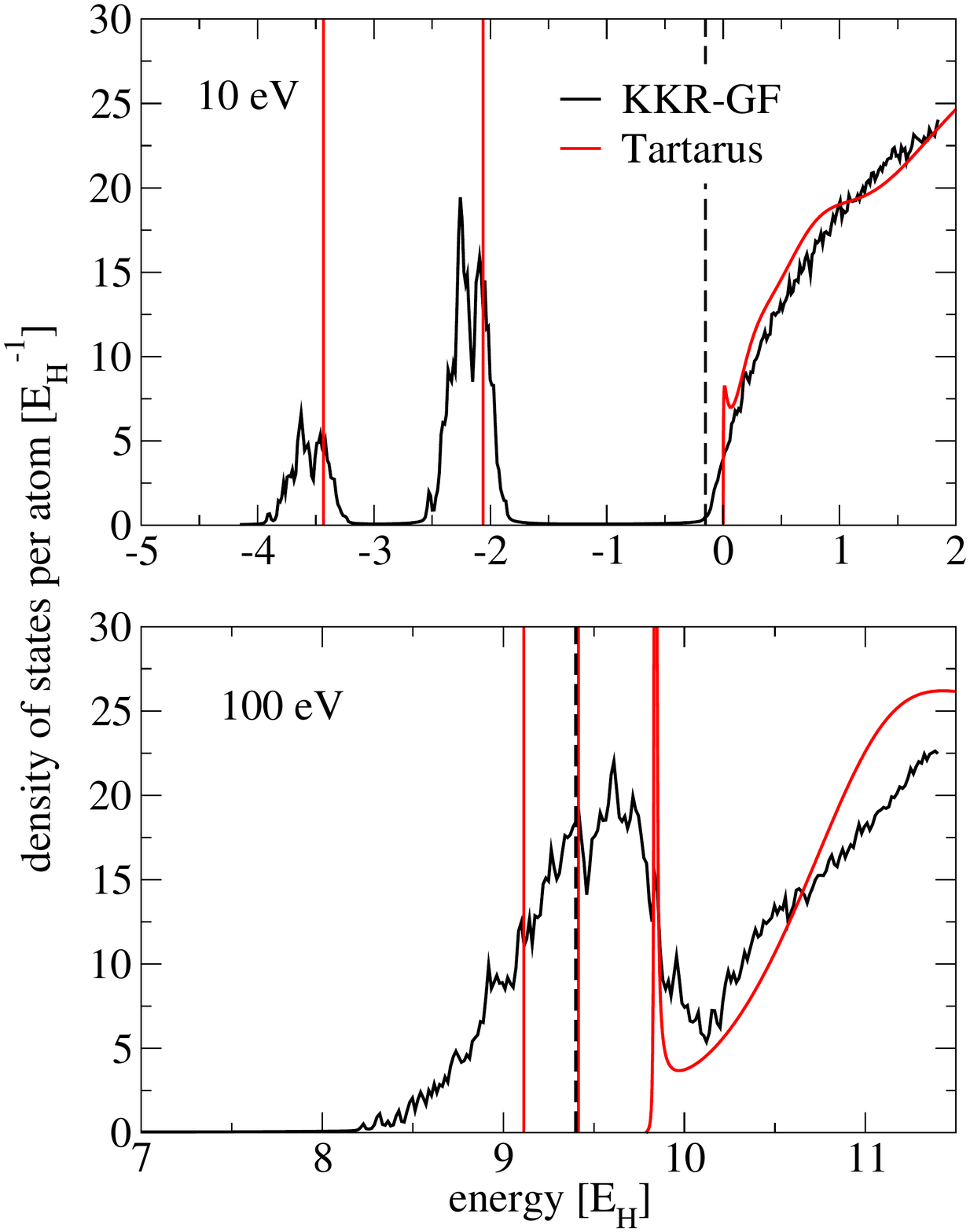}
\end{center}
\caption{(Color online) Density of states of solid density aluminum with $l_{max}^{ms}=2$ for temperatures of 10 eV (top panel) and 100 eV (bottom panel).   We also show the result from the average atom model \texttt{Tartarus} \cite{starrett19}.  We have made the zero of energy at the chemical potential.  For both cases the density of states shown is focused on the valence region, where states transition from bound to free.  The dashed vertical line shows the average energy of the muffin tin constant.
}
\label{fig_dos}
\end{figure}

\subsection{Diffraction and Disorder Effects on Electronic Structure}
In figure \ref{fig_dos} the density of states per atom is shown for the 10 and 100 eV aluminum cases, focusing on the valence region where multi-site, quantum diffraction effects are most important.  The density of states gives detailed information about the electronic structure, and is a less averaged quantity than the EOS.
To produce the density of states from the KKR-GF model, we have evaluated the Green's function slightly above the real energy axis.  This leads to a broadening of the the density of states \cite{johnson84}, equivalent to convolving with a Lorentzian of full-width-half-maximum equal to the imaginary part of the energy (here set to 0.1 E$_H$).  This is similar to the artificial broadening used in plane-wave DFT-MD to produce densities of states.

We compare to the average atom density of states using the {\texttt Tartarus} model, which is a spherically symmetric, single center model of an average atom in a jellium.  It has only one atom in it and hence no multiple scattering.  At 10 eV the {\texttt Tartarus} model yields two bound states in this energy range (2s and 2p), which are represented graphically by vertical lines corresponding to a Dirac-delta density of states.  In the KKR-GF model, these states broaden into bands of states, at roughly the same energy as the {\texttt Tartarus} bound states.  This broadening is the result of two effects.  First, since each atom has a different local environment, the eigenenergy varies from site to site.  Second, the inclusion of quantum diffraction through the multi-site term broadens the state, as was seen for high temperature ordered lattices \cite{starrett18}. The broadening can crudely be interpreted as a reduction in lifetime of the state due to multiple scattering, an effect that is absent in the average atom model.

For the average atom model, a distinct continuum in observed (figure \ref{fig_dos}) which corresponds to states with energies greater than the average atom muffin tin level.  These states are the charge carrying electron states in conductors.   At 10 eV, this continuum is also observed in the KKR-GF result and is more like the free electron form ($\propto \sqrt{\epsilon}$).  The artificially high symmetry in the average atom model leads to the structure in the continuum not seen in the KKR-GF model.

At 100 eV,  the average atom model predicts two bound states in the energy range shown (figure \ref{fig_dos}), as well as a resonance structure in the continuum. The resonance structure in average atom models is well documented \cite{starrett19,wilson06} and is a long lived quasi-stable state, caused by a potential minimum due to the sum of the effective potential and the centrifugal term $l(l+1)/r^2$ in the radial Kohn-Sham equation.  These bound and resonance states merge into one `bump' in the KKR-GF result, and the continuum merges with the bound states.  Resonances could occur in the KKR-GF model but are likely to be more spread out due to the fluctuations in the local enviroment and the multiple scattering effect.  This merging of the average atom bound and resonance states clearly has an influence on the EOS (see earlier).  Perhaps more dramatic would be the influence on optical spectra, as lines corresponding to transitions to and from these states in the average atom model would disappear and be replaced with some sort of merged feature.  This is the subject of future work.


\section{Conclusions and discussion}
The KKR-GF method has been demonstrated to be a viable and promising modeling tool for dense plasmas.  Calculations of equation of state demonstrate that the method gives good agreement with other state of the art methods, where they are valid and are practical.  Two major advantages of the method are 1) no pseudopotential is needed; core states are calculated explicitly and self-consistently, and 2) example calculations show that the computational time does not significantly change with temperature. 

There are limitations to our current implementation that are not limitations of the method in general.  First, we have used the muffin tin approximation.  This does not appear to cause major inaccuracies for the conditions explored in this work (i.e. temperature $>$ 10 eV), but we expect it to be more problematic at lower temperature where there only lower energy electrons, that will be more strongly affected by inaccuracies in the potential.  We note that the muffin-tin approximation is unnecessary, but serves us as a useful test vehicle.   Second, we have used a small number of atoms in the simulation cell (8), due to computer memory limitations, but a more sophisticated implementation could overcome this, possibly simulating thousands of atoms \cite{zeller11_book, wang95, thiess12}.  Third, we have imported the ion positions from a different model (PAMD \cite{starrett15}).  While this is accurate for the present results, molecular dynamics forces could be calculated directly from the electronic density, as is done in plane-wave DFT-MD.

A limitation of the method that has yet to be fully resolved is the convergence of the multi-site contribution to the Green's function.  Recent work has explored the possibility of transforming the conditionally convergent multi-site summation to an absolutely convergence one \cite{zeller15}, but to our knowledge, the practicality of this approach remains an open question.  We have used the empirically supported method of keeping only the chemically relevant expansion terms, and found only weak dependence of the pressure on the exact number, consistent with past results.

An exploration of the density of states of hot, dense aluminum plasmas reveals significant changes to the density of states compared to an average atom model.  There is broad alignment of features with this more approximate model, and the comparison facilitates understanding how multiple scattering affects the electronic structure.  It is pointed out that this will have significant impact on spectroscopic lines, for example, in the absorption coefficients.

\section*{Acknowledgments}
This work was performed under the auspices of the United States Department of Energy under contract DE-AC52-06NA25396.

\appendix

\section{Structural Green's function\label{app_struct}}
In this appendix we give the expression for the structural Green's function ${\cal G}^{nn'}_{LL'}$ for a periodic system \cite{segall57, treusch66}.  
The structural Green's function is first found in Fourier space using Ewald's technique and the transformed back to real space.  For the application presented here, we consider a periodic supercell in a simple cubic lattice.
Let $\mu$ be
the index the lattice vectors ${\bm a}_\mu$, and ${\bm R}_n$ be the vector pointing from the origin to the $n^{{th}}$ atom in the supercell.  
The real space structural Green's function
\begin{equation}
\begin{matrix}[2.0]
  {\cal G}_{LL^\prime}^{nn^\prime, \mu\mu^\prime} = & \frac{1}{V_{BZ}} \int_{BZ} d\bk
  e^{\imath \bk\cdot ({\bm R}_n - {\bm R}_{n^\prime})}
  e^{\imath \bk\cdot ({\bm a}_\mu - {\bm a}_{\mu^\prime})} \\
  & \times 
  \left[
  \left( 1 - {\bm {\mathcal G}}_0(\bk, z) {\bm t}(z)\right)^{-1} {\bm{\mathcal G}}_0(\bk,z)
  \right]_{LL^\prime}^{nn^\prime}
\end{matrix}
\end{equation}
where the matrices are indexed in $n$ and $L$, and the integral is over the Brillouin zone of volume $V_{BZ}$.  Note that, in the main text, the omission of the superscripts $\mu$ and $\mu'$  indicates that $\mu=\mu'$, referring to atoms in the same supercell.  The structural constants are given by
\begin{equation}
  {\cal G}_{0,LL^\prime}^{nn^\prime} = 2m\left \{ A_{lm,l^\prime m^\prime}^{nn^\prime} + 
                                  \imath p \delta_{ll^\prime}\delta_{mm^\prime}\delta_{nn^\prime} \right\}
\end{equation}
where
\begin{equation}
  A_{lm,l^\prime m^\prime}^{nn^\prime} = 4\pi \imath^{l-l^\prime} \sum_{l^{\prime \prime}}
  D_{l^{\prime\prime}, m-m^\prime}^{nn^\prime}
  C_{lm, l^{\prime\prime} m^{\prime\prime}}^{l^\prime m^\prime}
\end{equation}
The $C_{lm,\lpp\mpp}^{\lp\mmp}$ are called Gaunt coefficients
\begin{equation}
  C_{lm,\lpp\mpp}^{\lp\mmp}  \equiv
  \int d\hat{\bm r} 
  Y_{l,m}(\hat{{\bm r}})
  Y_{\lp,\mmp}(\hat{{\bm r}})^*
  Y_{\lpp,\mpp}(\hat{{\bm r}})
\end{equation}
and
\begin{equation}
  D_{l,m}^{nn^\prime} = D_{l,m}^{(1)nn^\prime} + D_{lm}^{(2)nn^\prime} +
   \delta_{nn^\prime}\delta_{l,0} \delta_{m,0} D_{0,0}^{(3)} 
\end{equation}
Here $D^{(1)nn'}_{l,m}$ is the part of Ewald's summation summed in k-space
\begin{equation}
\begin{matrix}[2.0]
  D_{l,m}^{(1)nn^\prime} = & - 4 \pi \exp{(z / \eta)}\frac{1}{V} p^{-l}
  \sum_{i} k_i^l (k_i^2-z)^{-1} \\
  & \times
  \exp(-k_i^2/\eta)Y_{lm}^*(\hat{\bk_i})
  e^{\imath \bk_i \cdot ({\bm R}_n - {\bm R}_{n^\prime})}
\end{matrix}
\end{equation}
with $\bk_i$ being a reciprocal space lattice vector,  the sum is over all these lattice vectors, and $V$ is the volume of the supercell.  $D^{(2)nn'}_{l,m}$ involves the real space summation,
\begin{equation}
\begin{matrix}[2.0]
  D_{l,m}^{(2)nn^\prime} = & (-2)^{l+1} \pi^{-\hf} p^{-l} 
  \sum_{s}^\prime [\imath^l \exp( \imath \bk\cdot {\bm a}_s)] \\
  & \times
  Y_{lm}^*(\widehat{{\bm a}_s - \bcr_{nn^\prime}}) |\ba_s-\bcr_{nn^\prime}|^l \\
  & \times
  \int_{\sqrt{\eta}/2}^{\infty} \xi^{2l} 
  \exp[ -\xi^2 (\ba_s-\bcr_{nn^\prime})^2 + (z/(4\xi^2))]d\xi
\end{matrix}
\end{equation}
where $\bcr_{nn^\prime} = \bcr_n - \bcr_{n^\prime}$ and the sum is over all lattice vectors, with the prime indicating that
the lattice vector ${\bm a}_s = {\bm 0}$ is ommitted.
\begin{equation}
  D_{0,0}^{(3)} = -\sqrt{\eta}(2\pi)^{-1} \sum_{j=0}^{\infty}
  (z/\eta)^j [ j! (2j-1) ]^{-1}
\end{equation}
$\eta$ is the Ewald parameter and the result is independent of its value (though
computation time is not \cite{bruno97}).
In implementing these equations, we found the work of reference \cite{davis71} helpful.

\bibliographystyle{unsrt}
\bibliography{phys_bib}

\end{document}